\documentclass[preprint,12pt]{elsarticle}

\usepackage{amssymb,amsmath}
\usepackage{breqn}
\usepackage{braket}
\usepackage{bm}
\usepackage{amsfonts}
\usepackage{hyperref}
\hypersetup{hidelinks}

\newcounter{bla}

\journal{Computer Physics Communications}

\begin{document}

\begin{frontmatter}



\title{Direct determination of layer anomalous Hall conductivity using uniaxial Wannier functions}


\author[a]{Yume Morishima\corref{author}}
\author[b]{Fumiyuki Ishii}
\author[b]{Naoya Yamaguchi\corref{author}}

\cortext[author] {Corresponding author.\\\textit{E-mail address:} morishima@cphys.s.kanazawa-u.ac.jp (Y. Morishima), \\ n-yamaguchi@cphys.s.kanazawa-u.ac.jp (N. Yamaguchi).}
\address[a]{Graduate School of Natural Science and Technology, Kanazawa University, Kakuma-machi, Kanazawa 920-1192, Japan}
\address[b]{Nanomaterials Research Institute (NanoMaRi), Kanazawa University, Kakuma-machi, Kanazawa 920-1192, Japan}

\begin{abstract}
We propose a method for computing layer anomalous Hall conductivity (LAHC) in real space by integrating the Fukui-Hatsugai-Suzuki method with hybrid Wannier functions localized along a single axis.
To validate the method, we calculated the LAHC of axion-insulating MnBi$_2$Te$_4$ and confirmed the agreement between the sum of LAHC on the surface and the surface AHC previously reported.
We further applied the method to antiferromagnetic Mn$_2$Bi$_2$Te$_5$ and examined the dependence on the magnetic structure of LAHC, identifying cases with and without axion insulating behavior.
This layer-resolved analysis offers a powerful tool for studying topological transport in complex materials, including heterostructures, and may guide the design of future devices based on the anomalous Hall effect with precise layer control.
\end{abstract}


\end{frontmatter}



\section{Introduction}
The anomalous Hall effect (AHE) is a phenomenon that an external electric field applied to a magnetic material produces a Hall current perpendicular to the electric field.
The anomalous Hall effect is also necessary for the emergence of the anomalous Nernst effect, a thermoelectric conversion phenomenon.
The AHE has an extrinsic mechanism induced by the impurity scatterings, such as skew scattering and side jump, and an intrinsic mechanism that originates from the Berry curvature associated with the band structure \cite{RevModPhys_AHC_Nagaosa,RevModPhys_AHC_Xiao}. 
Topological materials are typical materials with finite Berry curvatures, and studies of the AHE and quantized AHE in topological materials such as Weyl semimetals \cite{Weyl_Co3Sn2S2,Weyl_Co2MnGa,Weyl_Mn3Sn,Weyl_Mn3X}, nodal line semimetals \cite{nodal-line_Fe3GeTe2}, and Chern insulators have been conducted \cite{Cr-BiSbSe3}.
Berry curvature and Berry phase are closely related to topological properties in the electronic structure of materials, and they may be crucial to detailed electronic structure analysis to find guidelines for material design utilizing the anomalous Hall effect.


There have been several important introductions of the Berry phase \cite{BerryP} in computational theory in solid state physics \cite{TKNN}.
The description of electric polarization by the Berry phase was established early on as a fundamental calculation method in first-principles calculations based on density functional theory \cite{MTP,MTP_Resta}, which is now known as the ``modern theory of polarization.''
The “modern theory of polarization” demonstrated that the difference in electric polarization between two states of a solid crystal is a physical quantity of the bulk, as determined by the Berry phase.
When determining the Berry phase in solid crystals, it is necessary to calculate the overlap of the periodic parts of Bloch functions with adjacent wave numbers in the discretized reciprocal space.
Although the Bloch functions are delocalized in real space, a particle-like picture in the interpretation of electronic states can be obtained by using localized basis functions, i.e., Wannier functions, constructed by their unitary transformations \cite{PR_Wannier,RMP_Wannier}.
In fact, with the introduction of the Wannier function, the electric polarization is shown to be a constant multiple of the sum of the Wannier centers.
In other words, this suggests that electric polarization can be decomposed under the sum rule if the Wannier centers are obtained directly.

The simplest form of spatial decomposition is layer decomposition, which is decomposition in a uniaxial direction.
The hybrid Wannier function method is known as a uniaxial localization method \cite{PRB_HWF,HWF_review}.
Unlike conventional maximally localized Wannier functions, hybrid Wannier functions can be obtained directly as the most localized basis without iterative calculations.
Furthermore, based on hybrid Wannier functions, a definition of ``layer polarization'' corresponding to the contribution of each layer's electric polarization was derived \cite{PRL_LP}.
Thus, hybrid Wannier functions serve as a powerful tool for visualizing the layer decomposition of physical quantities.

The AHC in insulators can be calculated by the Fukui-Hatsugai-Suzuki method \cite{Fukui-Hatsugai-Suzuki}.
In this method, the Berry flux, which corresponds to the Berry curvature in the discretized FBZ, is calculated from the overlap matrices of the periodic parts of the Bloch functions between the occupied bands at adjacent wave number points.
This can be regarded as the Berry phase forming a locally closed circuit, and could be applied to metallic systems by considering it as a ``local Berry phase'' \cite{PRB_sawahata}.
Since the method for calculating the local Berry phase is equivalent to that used in the modern theory of polarization, spatial decomposition, especially layer decomposition, should be possible.
New measurement techniques such as magnetic imaging using measurement of local anomalous Nernst conductivity have also been proposed \cite{budai2023high,isshiki2023magneto,PhysRevLett.132.216702}.
Several theoretical studies have been conducted focusing on the locality of the Charn number and AHC in real space \cite{PhysRevB.84.241106,PhysRevB.95.121114,PhysRevB.98.245117}, and recently, calculations of layer-specific AHC have been performed for layered materials such as MnBi$_2$Te$_4$ \cite{MnBi2Te4_surfaceAHE,li2025high}.


In this paper, we propose an effective method for computing the AHC decomposed uniaxially in real space by combining the Fukui-Hatsugai-Suzuki method \cite{Fukui-Hatsugai-Suzuki} with hybrid Wannier functions \cite{PRB_HWF,HWF_review}, and introduce the “layer anomalous Hall conductivity” (LAHC).
The advantage of this method is that it does not require information on unoccupied states for the calculation.
To validate our method, we calculated the LAHC of MnBi$_2$Te$_4$, an axion insulator with an even number of septuple layers (SL), and confirmed that the sum of those belonging to one surface matched the surface AHC obtained in the previous study \cite{MnBi2Te4_surfaceAHE}.
We also calculated the LAHC of Mn$_2$Bi$_2$Te$_5$ and investigated the layer dependence for two antiferromagnetic magnetic structures, obtaining one that exhibits axion insulating behavior and the other that does not.

\section{Theory}
The anomalous Hall conductivity (AHC) $\sigma_{xy}$ induced by the intrinsic mechanism is expressed as the first Brillouin zone (FBZ) integral of the Berry curvature $\Omega_n(\bm{k}) \equiv -i\Bra{\nabla_{\bm{k}}u_{n\bm{k}}}\times\Ket{\nabla_{\bm{k}}u_{n\bm{k}}}$:
\begin{eqnarray}
\sigma_{xy} = \frac{e^2}{\hbar}\sum_{n}\int_{\mathrm{BZ}} \frac{d\bm{k}}{(2\pi)^d}f(\varepsilon_n(\bm{k}))\Omega_n^z(\bm{k}), \label{eq:1}
\end{eqnarray}
where $e, \hbar, d, f$, and $\varepsilon$ are the elementary charge, Dirac constant, dimension of the system, Fermi-Dirac distribution function, and band energy, respectively, and $\Ket{u_{n\bm{k}}}$ is the periodic part of Bloch states $\Ket{\psi_{n\bm{k}}} = e^{i\bm{k}\cdot\bm{r}}\Ket{u_{n\bm{k}}}$ with momentum space $\bm{k}$ and band index $n$ \cite{RevModPhys_AHC_Nagaosa,RevModPhys_AHC_Xiao}. 
As can be seen from Eq. (\ref{eq:1}), the $\sigma_{xy}$ generally does not contain information on the position in real space.

We address an insulating bulk system with electron occupation number denoted as $J$.
Let us introduce the wave vector $\bm{k}(I_a, I_b, I_c) = (I_a / N_a)\bm{G}_a + (I_b / N_b)\bm{G}_b + (I_c / N_c)\bm{G}_c$, where the number of $k$-points $N_a$, $N_b$, and $N_c$ discretize the first Brillouin zone along the reciprocal lattice vectors $\bm{G}_a$, $\bm{G}_b$, and $\bm{G}_c$, respectively.
We define the $J \times J$ overlap matrix $M_{\bm{k},\bm{k}+\delta\bm{k}_a}$ between $\bm{k}$ and  $\bm{k}+\delta\bm{k}_a (= \bm{k}(I_a + 1, I_b, I_c))$ points as
\begin{eqnarray}
\left(M_{\bm{k},\bm{k}+\delta\bm{k}_a}\right)_{mn} = \Braket{u_{m\bm{k}(I_a, I_b, I_c)} | u_{n\bm{k}(I_a + 1, I_b, I_c)}} \label{eq:2}
\end{eqnarray} 
and similarly define $M_{\bm{k}+\delta\bm{k}_a, \bm{k}+\delta\bm{k}_a + \delta\bm{k}_b}, M_{\bm{k}+\delta\bm{k}_a + \delta\bm{k}_b, \bm{k}+\delta\bm{k}_b}$ and $M_{\bm{k}+\delta\bm{k}_b, \bm{k}}$.
Suppose, for simplicity, that the $c$-axis of the crystal system is oriented along the $+z$ direction and is perpendicular to both the $a$-axis and the $b$-axis.
The AHC $\sigma_{xy}$ using the Fukui-Hatsugai-Suzuki method can be calculated as follows \cite{Fukui-Hatsugai-Suzuki} :
\begin{eqnarray}
F^J(\bm{k}(I_a, I_b, I_c)) &=& \mathrm{Im}\ln\left[\det(M_{\bm{k}, \bm{k}+\delta\bm{k}_a})\det(M_{\bm{k}+\delta\bm{k}_a, \bm{k}+\delta\bm{k}_a + \delta\bm{k}_b})\right.\nonumber\\
&\qquad& \quad \times \left.\det(M_{\bm{k}+\delta\bm{k}_a + \delta\bm{k}_b, \bm{k}+\delta\bm{k}_b})\det(M_{\bm{k}+\delta\bm{k}_b, \bm{k}})\right],  \label{eq:3}\\
\sigma_{xy} &=& \frac{e^2}{h}\frac{1}{2\pi}\frac{1}{N_c}\sum_{I_c = 0}^{N_c - 1}\sum_{I_a = 0}^{N_a - 1}\sum_{I_b = 0}^{N_b - 1} F^J(\bm{k}(I_a, I_b, I_c)). \label{eq:4}
\end{eqnarray}
Since the $\sigma_{xy}$ is defined at each $\bm{k}_c\left(\equiv(I_c / N_c)\bm{G}_c\right)$, the $\sigma_{xy}$ of bulk is determined by averaging those values at each $\bm{k}_c$.
The $F^J(\bm{k})$ is referred to as the Berry flux \cite{PRB_sawahata} and satisfies $-\pi \leq F^J(\bm{k}) < \pi$ because the operator $\mathrm{Im}\ln$ extracts the argument of the complex number \cite{Fukui-Hatsugai-Suzuki,PRB_sawahata}.

In our approach, we transform the expression for the AHC in the Fukui-Hatsugai-Suzuki method represented by Eqs. (\ref{eq:3}) and (\ref{eq:4}). 
This transformation involves converting the basis of the overlap matrix from the Bloch representation to the hybrid Wannier representation and enables the computation of the AHC decomposed along the one-axis in real space.
When we define the matrix product of the four overlap matrices as
\begin{eqnarray}
\tilde{M}(\bm{k}(I_a, I_b, I_c)) = M_{\bm{k}, \bm{k}+\delta\bm{k}_a}M_{\bm{k}+\delta\bm{k}_a, \bm{k}+\delta\bm{k}_a + \delta\bm{k}_b}M_{\bm{k}+\delta\bm{k}_a + \delta\bm{k}_b, \bm{k}+\delta\bm{k}_b}M_{\bm{k}+\delta\bm{k}_b, \bm{k}}, \label{eq:5}
\end{eqnarray}
the $\sigma_{xy}$ is rewritten as $\sigma_{xy} = (e^2/2\pi hN_c)\sum_{I_c, I_a, I_b} \mathrm{Im}\ln\det \tilde{M}(\bm{k}(I_a, I_b, I_c))$.
The determinant of the $J \times J$ matrix $\tilde{M}(\bm{k})$ is equal to the product of $J$ eigenvalues obtained by diagonalizing $\tilde{M}(\bm{k})$. Consequently, the $\sigma_{xy}$ can be further transformed as follows:
\begin{eqnarray}
\sigma_{xy} &=& \frac{e^2}{h}\frac{1}{2\pi}\frac{1}{N_c} \sum_{I_c = 0}^{N_c - 1}\sum_{I_a = 0}^{N_a - 1}\sum_{I_b = 0}^{N_b - 1}\mathrm{Im}\ln\prod_{i=1}^{J} \mu_i({\bm{k}}(I_a, I_b, I_c)), \nonumber\\
&=& \frac{e^2}{h}\frac{1}{2\pi}\frac{1}{N_c} \sum_{I_c = 0}^{N_c - 1}\sum_{I_a = 0}^{N_a - 1}\sum_{I_b = 0}^{N_b - 1}\sum_{i=1}^{J}\mathrm{Im}\ln \mu_i({\bm{k}}(I_a, I_b, I_c)), \label{eq:6}
\end{eqnarray}
Here, we denote the $i$th eigenvalue of the $\tilde{M}(\bm{k})$ as $\mu_i(\bm{k})$ and the corresponding eigenvector as $\Ket{a_{i\bm{k}}}$. 
Furthermore, we introduce a $J \times J$ diagonal matrix $\mu(\bm{k})$ with diagonal elements $\mu_i(\bm{k})$ and a $J \times J$ unitary matrix $V(\bm{k})$ which diagonalizes $\tilde{M}(\bm{k})$ such that $V_{mi}(\bm{k}) = \Braket{\psi_{m\bm{k}} | a_{i\bm{k}}}$.
Using this  $V(\bm{k})$, the eigenstates are described as $\Ket{a_{i\bm{k}}} = \sum_{m=1}^{J} V_{mi}(\bm{k})\Ket{\psi_{m\bm{k}}}$, representing a unitary transformation of Bloch states.
In short, the diagonalization of $\tilde{M}(\bm{k})$ is expressed as $V^{\dagger}(\bm{k})\tilde{M}(\bm{k})V(\bm{k}) = \mu(\bm{k})$.
At the current stage, the matrix $\tilde{M}(\bm{k})$ is represented in the Bloch basis, and there is no advantage in transforming $\sigma_{xy}$ from a determinant form to an eigenvalue form because Bloch functions are delocalized. However, by changing the basis of $\tilde{M}(\bm{k})$ to a localized basis in real space, it becomes possible to decompose $\sigma_{xy}$ into contributions from each localized state.
We define the diagonal matrix $\mathcal{F}(\bm{k})$ with diagonal elements given by $\mathrm{Im}\ln \mu_i({\bm{k}})$, and then we can naturally introduce the operator $\hat{\mathcal{F}}$. 
We refer to them as the Berry flux matrix and Berry flux operator, respectively. The $\sigma_{xy}$ is also given by the trase form as 
\begin{eqnarray}
\sigma_{xy} &=& \frac{e^2}{h}\frac{1}{2\pi}\frac{1}{N_c}\sum_{I_c = 0}^{N_c - 1}\sum_{I_a = 0}^{N_a - 1}\sum_{I_b = 0}^{N_b - 1}\mathrm{Tr}[\mathcal{F}(\bm{k}(I_a, I_b, I_c))] \nonumber\\
&=& \frac{e^2}{h}\frac{1}{2\pi}\frac{1}{N_c}\sum_{I_c = 0}^{N_c - 1}\sum_{I_a = 0}^{N_a - 1}\sum_{I_b = 0}^{N_b - 1}\sum_{i=1}^{J}\Braket{a_{i\bm{k}(I_a, I_b, I_c)} | \hat{\mathcal{F}} | a_{i\bm{k}(I_a, I_b, I_c)}}, \nonumber
\end{eqnarray}
and the relation 
\begin{eqnarray}
\hat{\mathcal{F}}\Ket{a_{i\bm{k}}} = \mathrm{Im}\ln \mu_i(\bm{k})\Ket{a_{i\bm{k}}}. \nonumber
\end{eqnarray}
is satisfied. (See \ref{appendix:HWC})
Since the trace is invariant under the unitary transformation, we adopt hybrid Wannier states $\Ket{h_{j\bm{k}_{\parallel}}^{m}}$ localized in the $z$-direction in real space as a basis for taking the trace.
Here, $j$ is the hybrid Wannier index, $\bm{k}_{\parallel} = \bm{k}_{\parallel}(I_a, I_b)$ is the in-plane (delocalized directions) wave vector, and the index $m$ runs over unit cells $Rm$ along the $z$-axis \cite{HWF_review}.
Then, We obtain the $\sigma_{xy}$ using the form of the hybrid Wannier states:
\begin{eqnarray}
\sigma_{xy} &=& \frac{e^2}{h}\frac{1}{2\pi}\frac{1}{N_c} \sum_{I_a, I_b}\sum_{j, m}\Braket{h_{j\bm{k}_{\parallel}(I_a, I_b)}^{m} | \hat{\mathcal{F}} | h_{j\bm{k}_{\parallel}(I_a, I_b)}^{m}}, \label{eq:7} \\
&=& \frac{e^2}{h}\frac{1}{2\pi}\frac{1}{N_c} \sum_{I_a, I_b} \sum_{j, m}\sum_{I_c, i} \Braket{h_{j\bm{k}_{\parallel}(I_a, I_b)}^{m} | a_{i\bm{k}(I_a, I_b, I_c)}} \nonumber\\
&\qquad& \quad \times \Braket{a_{i\bm{k}(I_a, I_b, I_c)} | h_{j\bm{k}_{\parallel}(I_a, I_b)}^{m}}\mathrm{Im}\ln \mu_i(\bm{k}(I_a, I_b, I_c)). \label{eq:8}
\end{eqnarray}

In the following, we consider a slab system.
In an insulating slab system ($\bm{k}_c = \bm0$ and $m = 0$), Eq. (\ref{eq:8}) can be rewritten as
\begin{eqnarray}
\sigma_{xy} &=& \frac{e^2}{h}\frac{1}{2\pi}\sum_{I_a, I_b} \sum_{j}\sum_{i} \Braket{h_{j\bm{k}_{\parallel}(I_a, I_b)}^0 | a_{i\bm{k}_{\parallel}(I_a, I_b)}} \nonumber\\
&\qquad& \quad \times
\Braket{a_{i\bm{k}_{\parallel}(I_a, I_b)} | h_{j\bm{k}_{\parallel}(I_a, I_b)}^0} \mathrm{Im}\ln \mu_i(\bm{k}_{\parallel}(I_a, I_b)). \nonumber
\end{eqnarray}
The eigenvector $\Ket{a_{i\bm{k}_{\parallel}}}$ is expressed as a unitary transformation of the hybrid Wannier state yielding 
\begin{eqnarray}
\Ket{a_{i\bm{k}_{\parallel}}} = \sum_{s=1}^{J}Y_{si}(\bm{k}_{\parallel})\Ket{h_{s\bm{k}_{\parallel}}^{0}}. \label{eq:9}
\end{eqnarray}
Thus,  we obtain the final expression as 
\begin{eqnarray}
\sigma_{xy} = \frac{e^2}{h}\frac{1}{2\pi}\sum_{I_a = 0}^{N_a - 1}\sum_{I_b = 0}^{N_b - 1} \sum_{j=1}^{J}\sum_{i=1}^{J}\left|Y_{ji}(\bm{k}_{\parallel}(I_a, I_b))\right|^2 \mathrm{Im}\ln \mu_i(\bm{k}_{\parallel}(I_a, I_b)). \label{eq:10}
\end{eqnarray}
 The $\sum_{i=1}^{J} \left|Y_{ji}(\bm{k}_{\parallel})\right|^2 \mathrm{Im}\ln \mu_i(\bm{k}_{\parallel})$ in Eq. (\ref{eq:10}) corresponds to $\Braket{h_{j\bm{k}_{\parallel}}^0 | \hat{\mathcal{F}} | h_{j\bm{k}_{\parallel}}^0}$ in eq. (\ref{eq:7}).
The transformation from Bloch states to hybrid Wannier states is expressed using the unitary matrix $U(\bm{k}_{\parallel})$ as 
\begin{eqnarray}
\Ket{h_{j\bm{k}_{\parallel}}^{0}} = \sum_{n=1}^{J} U_{nj}(\bm{k}_{\parallel})\Ket{\psi_{n\bm{k}_{\parallel}}}. \label{eq:11} 
\end{eqnarray}
From the eqs. (\ref{eq:9}) and (\ref{eq:11}), $\Ket{a_{i\bm{k}_{\parallel}}} = \sum_{n=1}^{J} \Ket{\psi_{n\bm{k}_{\parallel}}}(U(\bm{k}_{\parallel})Y(\bm{k}_{\parallel}))_{ni}$ is obtained.
Hence, the following equations are satisfied:
\begin{eqnarray}
V(\bm{k}_{\parallel}) &=& U(\bm{k}_{\parallel})Y(\bm{k}_{\parallel}), \label{eq:12}\\
V^{\dagger}(\bm{k}_{\parallel})\tilde{M}(\bm{k}_{\parallel})V(\bm{k}_{\parallel}) &=& Y^{\dagger}(\bm{k}_{\parallel})U^{\dagger}(\bm{k}_{\parallel})\tilde{M}(\bm{k}_{\parallel})U(\bm{k}_{\parallel})Y(\bm{k}_{\parallel}), \nonumber \\
&=& \mu(\bm{k}_{\parallel}), \label{eq:13}
\end{eqnarray}
Equation (\ref{eq:13}) demonstrates that we can get the unitary matrix $Y(\bm{k}_{\parallel})$ by transforming the basis of $\tilde{M}(\bm{k}_{\parallel})$ from Bloch representation to hybrid Wannier representation $\left(U^{\dagger}(\bm{k}_{\parallel})\tilde{M}(\bm{k}_{\parallel})U(\bm{k}_{\parallel})\right)$ and then diagonalizing it.

To obtain maximally localized hybrid Wannier states, it is necessary to choose the matrix $U(\bm{k}_{\parallel})$ appropriately.
For this purpose, we use the method discussed in reference to form the ``parallel transport'' basis for Bloch functions.
In the slab system, we consider the overlap matrix $M_{\bm{k}_{\parallel}, \bm{k}_{\parallel}+\bm{G}_c}$ between $\bm{k}_{\parallel}$ and $\bm{k}_{\parallel} + \bm{G}_c$.
Diagonalizing $M_{\bm{k}_{\parallel}, \bm{k}_{\parallel}+\bm{G}_c}$ yields the unitary matrix $U(\bm{k}_{\parallel})$ and the hybrid Wannier center $\bar{z}_j(\bm{k}_{\parallel}) = \Braket{h_{j\bm{k}_{\parallel}}^0 | z | h_{j\bm{k}_{\parallel}}^0}$ corresponding to the hybrid Wannier state $\Ket{h_{j\bm{k}_{\parallel}}^0}$. 

\begin{figure}[htbp]
\begin{center}
\includegraphics*[scale=0.48]{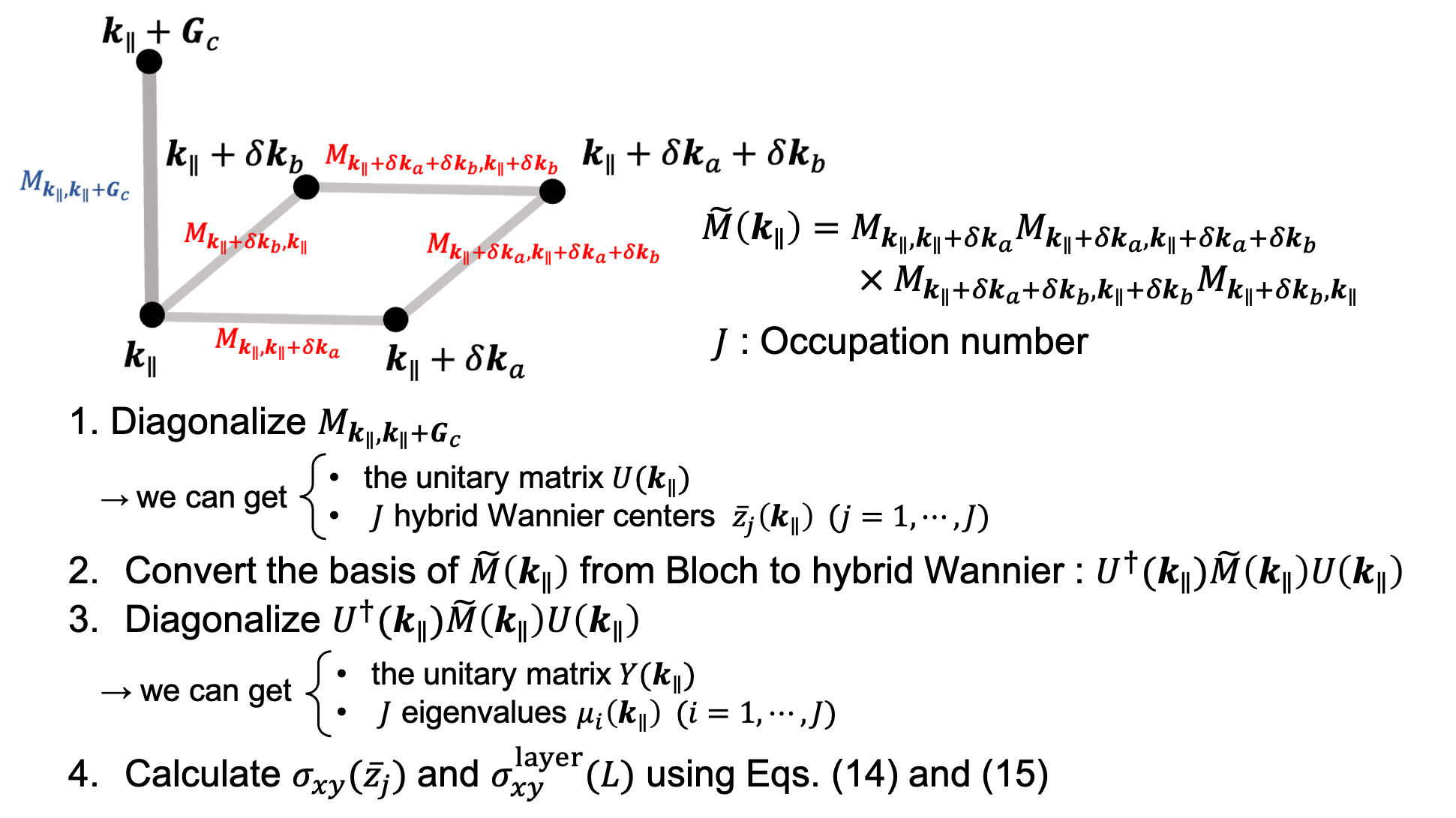}
\caption{The procedure of computing the $\sigma_{xy}(\bar{z}_j)$ in an insulating slab system.}
\label{fig:explanation}
\end{center}
\end{figure}

From Eq. (\ref{eq:10}), $\sigma_{xy}$ can be decomposed for each index $j$ of the hybrid Wannier state. 
Here, we denote the average of the $j$th hybrid Wannier center at all $\bm{k}_{\parallel}$ points $\sum_{\bm{k}_{\parallel}}\bar{z}_j(\bm{k}_{\parallel})/N_aN_b$ as $\bar{z}_j$.
We can define the $\sigma_{xy}$ at the hybrid Wannier center $\bar{z}_j$ as 
\begin{eqnarray}
\sigma_{xy}(\bar{z}_j) = \frac{e^2}{h}\frac{1}{2\pi}\sum_{I_a=0}^{N_a - 1}\sum_{I_b=0}^{N_b - 1} \sum_{i=1}^{J}\left|Y_{ji}(\bm{k}_{\parallel}(I_a, I_b))\right|^2 \mathrm{Im}\ln \mu_i(\bm{k}_{\parallel}(I_a, I_b)). \label{eq:14}
\end{eqnarray}
Furthermore, by taking the sum over $j$ where hybrid Wannier centers belong to layer index $L$, the LAHC
\begin{eqnarray}
\sigma_{xy}^{\mathrm{layer}}(L) = \sum_{\bar{z}_j \in L} \sigma_{xy}(\bar{z}_j) \label{eq:15}
\end{eqnarray}
can be obtained.
Note that the decomposition of AHC into layers can be defined in the same way as in “layer polarization,” and that LAHC here has the same unit as the original AHC, whereas “layer polarization” was the amount per unit area.
Figure \ref{fig:explanation} illustrates the procedure for calculating $\sigma_{xy}(\bar{z}_j)$.
As mentioned earlier, diagonalizing $M_{\bm{k}_{\parallel}, \bm{k}_{\parallel}+\bm{G}_c}$ allows obtaining the unitary matrix $U(\bm{k}_{\parallel})$, which enables the unitary transformation from Bloch states to hybrid Wannier states, and $J$ hybrid Wannier centers.
Using the unitary natrix $U(\bm{k}_{\parallel})$, we transform the basis of $\tilde{M}(\bm{k}_{\parallel})$ from Bloch to hybrid Wannier ( = $U^{\dagger}(\bm{k}_{\parallel})\tilde{M}(\bm{k}_{\parallel})U(\bm{k}_{\parallel})$) and diagonalize $U^{\dagger}(\bm{k}_{\parallel})\tilde{M}(\bm{k}_{\parallel})U(\bm{k}_{\parallel})$, yielding the unitary matrix $Y(\bm{k}_{\parallel})$ and $J$ eigenvalues $\mu_{i}(\bm{k}_{\parallel})$ ($i = 1, \cdots, J$).
Then, $\sigma_{xy}(\bar{z}_j)$ and LAHC $\sigma^{\mathrm{layer}}_{xy}(L)$ can be calculated using $Y(\bm{k}_{\parallel})$ and $\mu_i(\bm{k}_{\parallel})$ from Eqs. (\ref{eq:14}) and (\ref{eq:15}).

The advantages of this method are that it does not require information on unoccupied bands, as in the Fukui-Hatsugai-Suzuki method, and that the process of maximizing the localization of hybrid Wannier functions is performed by a direct method that does not require iterative calculations.

\section{Models}

\begin{figure}[htbp]
\begin{center}
\includegraphics*[scale=0.6]{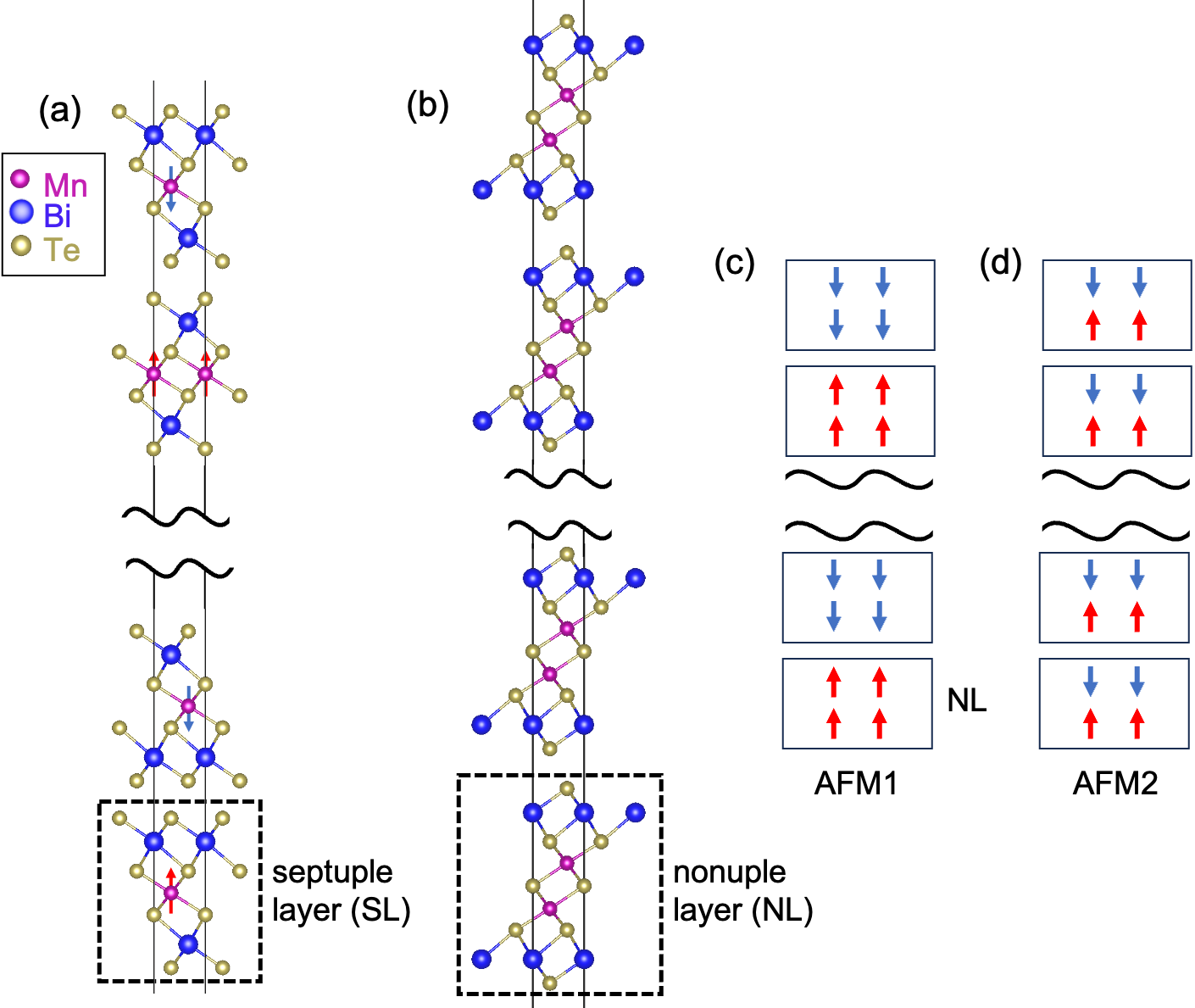}
\caption{Crystal structures of MnBi$_{2}$Te$_{4}$ with 10SLs and Mn$_{2}$Bi$_{2}$Te$_{5}$ with 10NLs. (a)MnBi$_{2}$Te$_{4}$. The red upward and blue downward arrows show the up and down spins of each Mn atom, respectively. The intralayer exhibits ferromagnetic order, while the interlayer exhibits antiferromagnetic order. (b)Mn$_{2}$Bi$_{2}$Te$_{5}$. (c), (d)The spin alignment of Mn$_{2}$Bi$_{2}$Te$_{5}$. We denote the magnetic orders of (c) and (d) as AFM1 and AFM2, respectively.}
\label{fig:structure}
\end{center}
\end{figure}

In axion insulators, one of the topological materials, the AHC at the top and bottom surfaces were predicted to be quantized to $+e^2/(2h)$ or $-e^2/(2h)$, respectively, depending on the magnetic orientation \cite{PRB_AFMTI}.
The computational models used in this study are shown in Fig. \ref{fig:structure}.
MnBi$_{2}$Te$_{4}$ is a van der Waals layered compound consisting of a stacking sequence of Te-Bi-Te-Mn-Te-Bi-Te septuple layers (SLs) along the $z$-axis (Fig. \ref{fig:structure}(a)).
It exhibits A-type antiferromagnetic order with intralayer ferromagnetic coupling within each SL and interlayer antiferromagnetic coupling between neighboring SLs in the ground state.
In the same way, two antiferromagnetic magnetic orders were considered for Mn$_2$Bi$_2$Te$_5$ because each SL contains two Mn atoms within the unit cell (Fig. \ref{fig:structure}(b-d)).

\section{Computational details}
We conducted first-principles calculations based on density functional theory (DFT) to obtain electronic structures and subsequently calculated the AHC decomposed along the $z$-axis in real space and resolved by layer, .i.e., LAHC.
For the self-consistent field calculations to obtain the electronic structures, we used the OpenMX with norm-conserving pseudopotentials and wave functions expanded by the linear combination of the pseudoatomic orbitals \cite{PhysRevB.67.155108,PhysRevB.69.195113,PhysRevB.72.045121}.
The LAHC calculations were performed using our implementation \cite{layerAHC}.
The Perdew–Burke–Ernzerhof generalized gradient approximation was employed for the exchange-correlation term \cite{PhysRevLett.77.3865}.
To take the localization of 3$d$ orbitals of Mn into account, the Hubbard $U$ method was used with $U_{\mathrm{eff}} = 4.0$ eV for Mn based on the previous study.
A non-collinear spin density functional with two-component spinor wavefunctions was used \cite{von1972local,kubler1988density}.
Spin–orbit coupling was considered through the
fully relativistic total-angular-momentum-dependent
pseudopotentials \cite{PhysRevB.64.073106}.
The norm-conserving pseudopotentials included the 3$s$, 3$p$, 3$d$, and 4$s$ electrons for Mn as valence electrons, while the 5$d$, 6$s$, and 6$p$ for Bi; 4$d$, 5$s$, and 5$p$ for Te were also considered.
We used a cutoff energy of 300 Ry and employed a $k$-point grid of $9 \times 9\times 1$. 
The pseudoatomic orbital basis sets were assigned as Mn6.0-$s3p2d2f1$, Bi8.0-$s3p3d2f1$, and Te7.0-$s3p3d2f1$. 
Here, the numerical value following each element denotes the radial cutoff in units of Bohr, and the integers after $s$, $p$, $d$, and $f$ represent the numbers of $s$-, $p$-, $d$-, and $f$-orbital sets, respectively.
In the calculation of AHC $\sigma_{xy}$ decomposed along the $z$-axis, a $100\times 100\times 1$ $k$-point grid was used for calculations at the Fermi energy.
The overlap matrix $M_{\bm{k},\bm{k}+\delta\bm{k}_a}$, which could be a bottleneck in the calculation, was computed by transforming it into the product of three matrices found in our previous study \cite{yamaguchi2022first}.

\section{Results and discussion}
\begin{figure}[htbp]
\begin{center}
\includegraphics*[scale=0.4]{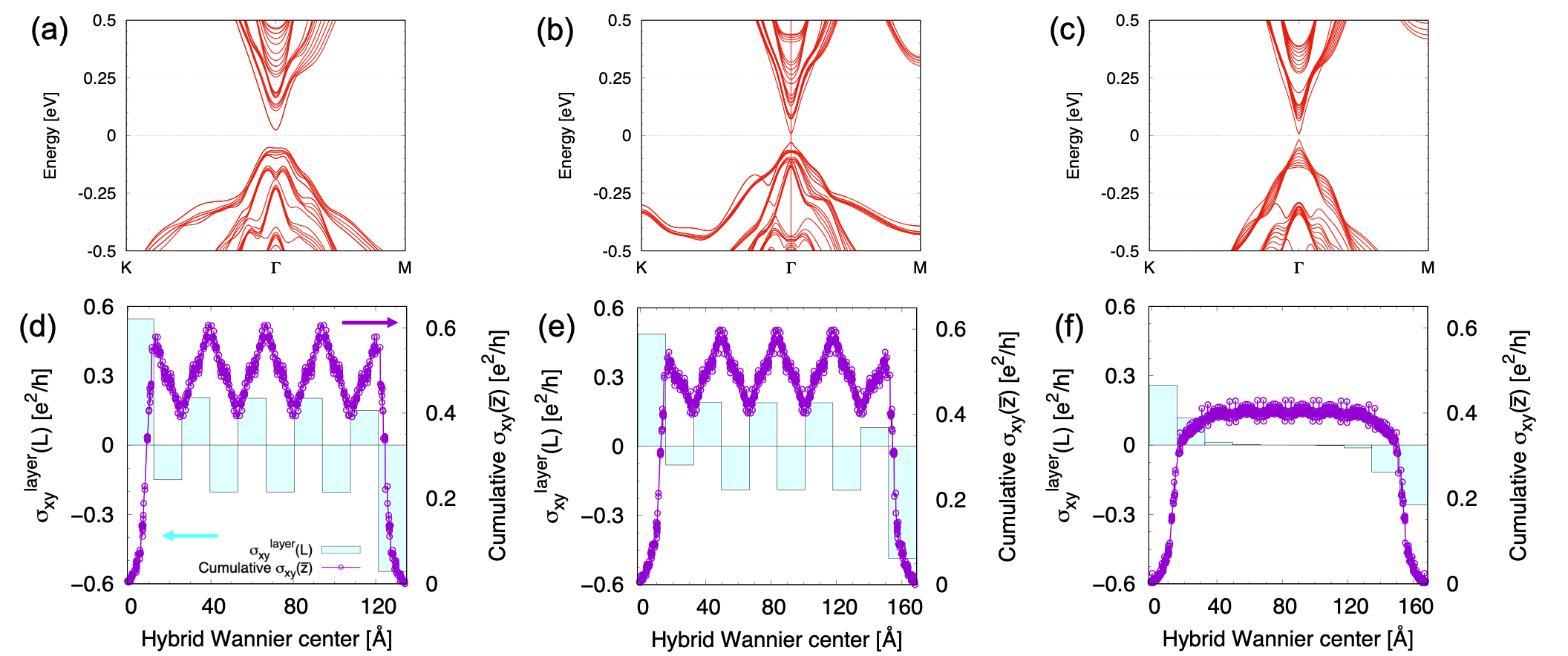}
\caption{(a-c) Band structures of (a) MnBi$_{2}$Te$_{4}$ with 10SLs, (b) AFM1 Mn$_{2}$Bi$_{2}$Te$_{5}$ with 10NLs, and (c) AFM2 Mn$_{2}$Bi$_{2}$Te$_{5}$ with 10NLs. (d-f) Layer-resolved $\sigma_{xy}^{\mathrm{layer}} (L)$ and Cumulative $\sigma_{xy}(\bar{z})$ at the Fermi energy. (d) MnBi$_{2}$Te$_{4}$ with 10SLs. The top and bottom SL contribute $+0.55$ $e^2/h$ and $-0.55$ $e^2/h$ in the $\sigma^{\mathrm{layer}}_{xy}$, respectively, while the $\sigma_{xy}^{\mathrm{layer}}$ oscillate in the intermediate SLs due to changes in magnetic ordering, resulting in the total $\sigma_{xy}$ of 0. (e) AFM1 Mn$_{2}$Bi$_{2}$Te$_{5}$ with 10NLs. Similar to the case of MnBi$_{2}$Te$_{4}$, $\sigma^{\mathrm{layer}}_{xy}$ reached $+0.49 e^2/h$ and $-0.49 e^2/h$ in the top and bottom SL, respectively, and oscillated in the intermediate SLs. (f) AFM2 Mn$_{2}$Bi$_{2}$Te$_{5}$ with 10NLs. $\sigma^{\mathrm{layer}}_{xy}(1) = 0.26 e^2/h$ was observed, while the sum of the $\sigma^{\mathrm{layer}}_{xy}$ from 1 to 5SL reached to $0.39 e^2/h$, yielding a surface anomalous Hall conductivity that is not quantized to $\sim e^2/(2h)$.}
\label{fig:AHC}
\end{center}
\end{figure}

First, we verified the validity of our calculation method using an antiferromagnetic slab of MnBi$_2$Te$_4$ with 10SLs.
The bulk phase of MnBi$_2$Te$_4$ has been predicted to be an axion insulator and exhibit surface anomalous Hall conductivity quantized to $\pm e^2/(2h)$.
Figure \ref{fig:AHC}(a) represents the electronic structure near the Fermi energy, and the band gap was opened with the magnitude of 75 meV due to the magnetic ordering.
The LAHC $\sigma^{\mathrm{layer}}_{xy}(L)$ in each SL and the cumulative AHC represented by $\sum_{j^{\prime}=0}^{j} \sigma_{xy}(\bar{z}_{j^{\prime}})$ are shown in Fig. \ref{fig:AHC}(d).
At the bottom SL, $\sigma^{\mathrm{layer}}_{xy}(1)$ reached $0.55 e^2/h$, and the sum of $\sigma^{\mathrm{layer}}_{xy}(L)$ from 2 to 9 SL oscillated with the amplitude of $0.20 e^2/h$ due to changes in magnetic ordering. At the top SL, $\sigma^{\mathrm{layer}}_{xy}(10)$ achieved $-0.55 e^2/h$, resulting in an overall $\sigma_{xy} = 0$.
In the previous study on the antiferromagnetic MnBi$_2$Te$_4$ with 16SLs, $\sigma_{xy}(1)$ and $\sigma_{xy}(16)$ were $+0.49 e^2/h$ and $-0.49 e^2/h$ respectively, and the sum of $\sigma^{\mathrm{layer}}_{xy}(L)$ from 2 to 15SL exhibited oscillations with the amplitude of $0.21 e^2/h$ \cite{MnBi2Te4_surfaceAHE}.
Thus, the effectiveness of our computational method was confirmed because our result exhibited a trend similar to that of the previous study, and the surface AHC with $\sim \pm e^2/(2h)$ were achieved.

For Mn$_{2}$Bi$_{2}$Te$_{5}$, we also calculated the LAHC $\sigma^{\mathrm{layer}}_{xy}(L)$ along the $z$-axis assuming two magnetic structures in a slab with 10NLs.
As shown in Figs. \ref{fig:AHC}(b) and \ref{fig:AHC}(c), the band gap of 23 meV and 35 meV appeared for interlayer antiferromagnetism and intra-layer antiferromagnetism, i.e., AFM1 and AFM2, respectively.
The overall AHC $\sigma_{xy}$ vanished and the layer dependence of LAHC $\sigma^{\mathrm{layer}}_{xy}(L)$ was visualized as shown in Figs. \ref{fig:AHC}(e) and \ref{fig:AHC}(f).
In the case of interlayer antiferromagnetism, i.e., AFM1, oscillations similar to those found in MnBi2Te4 were observed, exhibiting axion insulator-like behavior, and the surface AHC was estimated to be $\pm0.49 e^2/h$.
However, in the case of intra-layer antiferromagnetism, i.e., AFM2, oscillation was suppressed, and surface AHC appeared, with a value of $\pm0.39 e^2/h$.
Despite the overall AHC being zero, as shown above, the LAHC profile and surface AHC values differed greatly due to differences in magnetic structure.
In other words, the LAHC profile is suggested to reflect not only magnetic but also topological properties, and visualization of LAHC is expected to be a new analytical tool useful for investigating such properties.

\section{Summary}
We developed a method for calculating layer-decomposed anomalous Hall conductivity by combining the Fukui-Hatsugai-Suzuki method, an effective technique for calculating $\sigma_{xy}$ in insulators, with hybrid Wannier functions localized in the real space along a single axis, and proposed the concept of “layer anomalous Hall conductivity (LAHC).”
We calculated the LAHC of MnBi$_2$Te$_4$ and Mn$_2$Bi$_2$Te$_5$ and evaluated the surface AHC values. When the surface AHC is approximately $e^2/(2h)$, as in the case of axion insulators, the LAHC profile oscillates, and when it is not, the oscillation is suppressed.
We found that, depending on the magnetic structure, there are cases where the value is not close to $e^2/(2h)$ and has a finite surface AHC.
In recent years, the AHE has been extensively reported in heterostructures, twisted systems, and amorphous materials \cite{zhou2022heterodimensional,yoo2021large,deng2021high,twist_AHE,PhysRevMaterials.4.114405}.
In these materials, an analysis of the LAHC in real space would yield a more profound comprehension.
Analysis of LAHC may serve as a tool to promote understanding of topological electronic transport properties in complex materials, establishing a foundation for analyzing various magnetic topological insulators and magnetic metal thin films.
It is also expected to provide useful information for device development utilizing the anomalous Hall effect, particularly when considering new material designs involving precise layer control, such as artificial superlattices.

\section*{Acknowledgment}
This work was supported by JSPS KAKENHI Grant Numbers JP20K15115, JP22H05452, JP22K04862, JP23H01129, JP23K23157, JP24K02950, JP25H01525, JP25K22797 and JST SICORP Program Grant Number JPMJSC21E3.
This work was also supported by Kanazawa University SAKIGAKE project.
The computation in this work has been done using the facilities of the Supercomputer Center, the Institute for Solid State Physics, the University of Tokyo.
Crystal structures were drawn by VESTA \cite{Momma2011}




\appendix
\section{Evaluation of hybrid Wannier centers}
\label{appendix:HWC}
We perform singular value decomposition $M_{\bm{k}_{\parallel}, \bm{k}_{\parallel}+\bm{b}_3} = X\Sigma W^{\dagger}$, where $X$ and $W$ are unitary matrices and $\Sigma$ is the diagonal matrix with nonnegative diagonal elements, and obtain the optimal unitary matrix $U(\bm{k}_{\parallel})$ and hybrid Wannier centers by diagonalizing the unitary part of the $M_{\bm{k}_{\parallel}, \bm{k}_{\parallel}+\bm{b}_3}$ denoted as $\mathcal{M}_{\bm{k}_{\parallel}, \bm{k}_{\parallel}+\bm{b}_3} \left( = XW^{\dagger}\right)$:
\begin{eqnarray}
U^{\dagger}(\bm{k}_{\parallel})\mathcal{M}_{\bm{k}_{\parallel}, \bm{k}_{\parallel}+\bm{b}_3}U(\bm{k}_{\parallel}) &=& \lambda(\bm{k}_{\parallel}), \label{eq:A-1} \\
\bar{z}_j(\bm{k}_{\parallel}) &=& -\frac{c}{2\pi}\mathrm{Im}\ln \lambda_j(\bm{k}_{\parallel}), \label{eq:A-2}
\end{eqnarray}
where $\lambda(\bm{k}_{\parallel})$ is the diagonal matrix with diagonal elements representing the eigenvalues $\lambda_j(\bm{k}_{\parallel})$, and $\bar{z}_j(\bm{k}_{\parallel}) \left(= \Braket{h_{j\bm{k}_{\parallel}}^0 | z | h_{j\bm{k}_{\parallel}}^0}\right)$ and $c$ represent the $j$th hybrid Wannier center and the lattice constant, respectively.

\bibliographystyle{jpsj}
\bibliography{reference}







\end{document}